\documentclass{aa}

\def\v2c{v_{\rm c}^2}

\begin{document}
\input{epsf}
\newbox\grsign \setbox\grsign=\hbox{$>$} \newdimen\grdimen \grdimen=\ht\grsign
\newbox\simlessbox \newbox\simgreatbox
\setbox\simgreatbox=\hbox{\raise.5ex\hbox{$>$}\llap
     {\lower.5ex\hbox{$\sim$}}}\ht1=\grdimen\dp1=0pt
\setbox\simlessbox=\hbox{\raise.5ex\hbox{$<$}\llap
    {\lower.5ex\hbox{$\sim$}}}\ht2=\grdimen\dp2=0pt
\def\simgreat{\mathrel{\copy\simgreatbox}}
\def\simless{\mathrel{\copy\simlessbox}}
\newbox\simppropto
\setbox\simppropto=\hbox{\raise.5ex\hbox{$\sim$}\llap
{\lower.5ex\hbox{$\propto$}}}\ht2=\grdimen\dp2=0pt
\def\simpropto{\mathrel{\copy\simppropto}}

\title{ISM gas removal from starburst galaxies and the premature death of star
clusters}
\author{
Claudio Melioli\inst{1} \and Elisabete M. de Gouveia Dal Pino
\inst{1} }
\offprints{Melioli C.}
\institute{ Universidade de S\~ao Paulo, IAG, Rua do Mat\~ao 1226,
Cidade Universit\'aria, S\~ao Paulo 05508-900, Brazil\\
e-mail: cmelioli@astro.iag.usp.br , \ \ dalpino@astro.iag.usp.br }

\date{Received ; accepted}

\abstract{Recent observational studies of the age distribution of
star clusters in nearby merging galaxies and starburst (SB)
galaxies  indicate a premature death of the young clusters.
The fate of an evolving star cluster crucially depends of
its gas content. This behaves like a  glue that helps to keep the
star system gravitationally bound. In SB systems where the rate of
supernovae (SNe) explosions is elevated one should  expect an
efficient heating of the gas and its complete removal which could
then favor the rapid dissociation of the evolving star clusters.
Based on a contemporaneous study of the dynamical evolution of the
interstellar gas in SB environments (Melioli \& de Gouveia Dal
Pino 2004) where
it has been considered also
the presence of dense clouds that may inhibit the heating
efficiency of the interstellar gas by the SNe, we  have here
computed the  timescales  for gas removal from  young clusters
embedded in these systems and found that they are consistent with
the very short timescales  for cluster dissolution which are
inferred from the observational studies above. Our results
indicate that typical SB proto-clusters should start to disperse
after less than  5 Myr. For a given total gas mass content, this
result is nearly insensitive to the initial star formation
efficiency.

}

\titlerunning{On the premature death of star clusters}

\authorrunning{Melioli \& de Gouveia Dal Pino}

\maketitle

\section{Introduction}

Star clusters in nearby SB galaxies seem to contribute with only
$\sim$20\% of the light seen in UV images, so that the diffuse
light is dominated by the contribution of field stars (Meurer et
al. 1995).
There is increasing observational evidence that suggests that this
diffuse UV field light is actually due to the dispersion of aged
star clusters that have left their remaining stars in the field
(Tremonti et al. 2001, Lada \& Lada 2003, Fall 2004). Chandar et
al. (2005) have observed this trend in eleven nearby SB galaxies
and have also identified a lack of massive stars in the field
which supports the interpretation above that the field stars are
composed of older, dissolving clusters. They also estimate that
the star clusters need to dissolve on timescales 7-10 Myr in order
to create the field stars.

A contemporaneous statistical study of the age distribution of
star clusters in nearby merging galaxies (the Antannae galaxies)
by Fall (2004; see also Whitmore et al. 2005) has shown that the
number of clusters decreases by a factor of $\sim$ 10 by the time
the cluster population has reached the age of $\sim$ 10 Myr. This
rapid decline is very insensitive to the cluster mass, at least
for masses above 3 $\times 10^4$ solar masses,
and
should be an indication that most of the clusters have become
gravitationally unbounded.

The fraction of gas mass that is converted into stars in SBs is
about 10 times larger than in normal galaxies and, as such, the
production of a high number of massive stars results high rates of
SN explosions. Thus SN explosions, as well as the ionizing
radiation from massive stars, and stellar winds may efficiently
heat the gas and eventually remove a significant fraction of the
ISM from a proto-cluster, leaving the stars gravitationally
unbound. The crossing time for stars orbiting within a bound
cluster or proto-cluster is $\tau_{cr} \simeq 2 R/ \sigma$, where
$\sigma$ is the velocity dispersion of cluster,
$\sigma \simeq [0.4 G(M_{star} + M_{gas})/R]^{1/2}$,
and $R$ is its the radius.
Now, if a proto-cluster suddenly loses most of
its mass by the removal of the gas and the instantaneous velocity
dispersion of the stars at the time of gas removal, $\sigma$,
becomes larger than the escape velocity,
$v_{esc} \simeq [(2GM_{star})/R]^{1/2}$,
then the cluster will be no longer gravitationally bound and will
expand almost freely with its radius increasing with age as (see,
e.g., Fall 2004; Lada \& Lada 2003) $R_{\tau} \simeq R (\tau/
\tau_{cr})$ and its surface density decreasing as $\Sigma_{\tau}
\simeq \Sigma (\tau / \tau_{cr})$.

A recent independent theoretical study of the evolution of the ISM
of SB environments by Melioli \& de Gouveia Dal Pino (2004; see
also Melioli, de Gouveia Dal Pino \& Raga 2005), has revealed that
the effectiveness of the SNe and their shock waves to heat the IS
gas and make it to be ejected from the SB system is sensitive to
the total gas mass content. In particular, we have found that if
the total gas mass both in the form of clouds and diffuse gas is
large enough, then the radiative cooling time scale of the gas can
remain shorter than the timescale for the development of a hot
superbubble (due to the SNRs interactions) for about half lifetime
of the SB.
During this time, most of the gas is retained in the SB, but after
this period the SNe heating efficiency rapidly increases to its
maximum value thus leading to gas expansion and removal from the
SB in a short period.
The clouds play a fundamental
role in this process as they work like valves that are able to
both  retain part of the gas to themselves and  lose part of it to
the diffuse ISM through cloud photoevaporation and then, maintain
the ISM density and the radiative cooling rate high enough for a
long time. As long as the gas is retained it may favor new
generations of star formation.

In this work, we apply the analysis above, which was performed for
an entire SB system (Melioli \& de Gouveia Dal Pino2004), to the
scale of individual star proto-clusters embedded in it, and show
that the resulting typical timescales both for gas retention and
removal from these clusters are consistent with the very short
timescales for cluster dissolution inferred from the previous
investigations of Fall (2004) and Chandar et al. (2005) of local
SB galaxies

\section{A Model for the Evolution of the gas}
At least two physical parameters are essential to determine the
evolution of a proto-cluster: the star formation efficiency and
the timescale of gas dispersal from the cluster (e.g., Lada \&
Lada 2003).

Accurate measurements of the star formation efficiency, SFE =
$M_{star}/ (M_{gas} + M_{star})$, which require a reliable
determination of both the gaseous and the stellar mass content,
are in general not available for cluster forming regions, nor even
in our galaxy. In the latter, the inferred SFEs of proto-clusters
embedded in giant molecular clouds range from $\sim$ 10\% to 30\%.

The timescale for gas removal, $\tau_g$, is even less constrained
by empirical data.
In SB systems, in particular where the rates of
SN explosions are elevated, one should expect an efficient heating
of the gas followed by its complete removal from the
proto-cluster. The condition for a cluster to become
gravitationally unbound after a rapid gas dispersal (e.g., in a
timescale $\tau_g \le \tau_{cr}$) is that the instantaneous
velocity  dispersion  of the stars  at the time of  gas removal is
larger than the escape velocity, or $\sigma > v_{esc}$. From the
equations for $\sigma$ and  $v_{esc}$ given in Section 1, one
finds that this condition implies that a system with SFE $\le$
50\% should become unbound after rapid gas removal.
(e.g., Lada \& Lada 2003).
In the opposite case of a slow gas removal ($\tau_g < \tau_{cr}$)
even clusters with low SFEs would have time enough to adjust
adiabatically and expand to new states of virial equilibrium and
remain bound.

However, the simple estimates above,  do not take into account the
fact that proto-clusters are actually far more complex systems.
For instance, the gas is not homogeneously distributed but may be
mostly within dense small clouds.
These clouds will suffer mass loss by ablation, thermal
evaporation and photo-evaporation due to the large amount of
ionizing photons emitted by the massive stars and by the SNe
(e.g., Melioli \& de Gouveia Dal Pino 2004, 2005). The diffuse gas
will be  heated by the SNe, but may cool via radiative losses
which depend on the square of the gas density, so that on one hand
the SN remnant (SNR) shock fronts may cause the heating and
expansion of the diffuse gas, but on the other hand, the mass loss
from the clouds may increase the density of the diffuse gas enough
to allow an efficient radiative cooling that may delay the gas
removal.
Defining the heating efficiency of the SNe, HE, as the fraction of
their energy that is effectively stored by the gas in the form of
enthalpy and is not radiated away, then one may expect that an
efficient gas expansion and removal will occur only when HE
becomes close to unity.

In order to compute the evolution of the gas of a proto-cluster
embedded in a SB environment considering the different physical
processes mentioned above, we will adopt the semi-analytical
dynamical model described by Melioli \& de Gouveia Dal Pino (2004,
hereafter MGDP).

As in MGDP, we assume that an instantaneous burst of star
formation occurs inside a spherical parent cloud of radius
$R_{\rm}$. We adopt a Salpeter initial mass function (IMF) with a
number  of stars with mass greater than 8 M$_{\odot}$ (i.e., the
SN number) ${\cal N} \sim 0.01 M_{star}$, where $M_{star}$ is the
total mass of  stars in M$_{\odot}$.

For simplicity, it is assumed that the expanding SNRs and the
cold, dense cloud clumps are uniformly distributed in the parent
cloud. The clumps have an initial density $n_c \sim 10^4$, radius
$r_c \sim 0.02$ pc, and mass $M_c=0.015$ M$_{\odot}$.
Assuming similar characteristics to those observed in the Rosette
giant cloud in our galaxy,
in most of the calculations below we will adopt a total mass of
clumps
 $M_{gas} \simeq 8 M_{star}$.

The equations that describe the evolution of the gas are given in
MGDP. In the Section below, we will present  the evolution of the
diffuse gas in a proto-cluster for different values of the SFE
until the time when the heating efficiency, HE, which is
mathematically defined as the ratio of the enthalpy flux through
the boundary of the system to the total SN energy flux:

\begin{equation}
{\rm HE}= 4 \pi R^2 {{({1 \over 2}\rho v^2+{3 \over 2}p)}\over
{{\cal R} \rm E_{SN}}} {\cal M} c_s,
\end{equation}
\noindent becomes equal to unity.

\section{Results}
Table 1 summarizes the input parameters of the computed models.
The values adopted are compatible with those estimated from the
observations (e.g., Chandar et al. 2005, Fall 2004, Tenorio-Tagle
et al. 2005 and references therein).

\begin{table*}
{\small \centerline{
\begin{tabular}{|c|c|c|c|c|c|c|c|}
\hline\hline \noalign{\smallskip} \hbox{Model} & \hbox{$n$
(cm$^{-3}$)} & \hbox{T (K)} & \hbox{R (pc)} & \hbox{$M_{gas}$
(M$_{\odot}$)} & \hbox{$M_{star}$
(M$_{\odot}$)} \\
\noalign{\smallskip} \hline \noalign{\smallskip} \hbox{1} &
\hbox{10} & \hbox{$10^4$} & \hbox{10} & \hbox{$8 \times 10^6$} &\hbox{$10^6$} \\
\noalign{\smallskip} \hline \hbox{2} & \hbox{10} & \hbox{$10^4$}
&\hbox{20} & \hbox{$8 \times 10^6$} &
\hbox{$10^6$} \\
\noalign{\smallskip} \hline \hbox{3} & \hbox{10} & \hbox{$10^4$}
&\hbox{5} & \hbox{$8 \times 10^5$} &
\hbox{$10^5$} \\
\hline \hbox{4} & \hbox{10} & \hbox{$10^4$} & \hbox{10} & \hbox{$8
\times 10^5$} & \hbox{$10^5$} \\
\noalign{\smallskip} \hline
\end{tabular}}
}
\end{table*}

Figure 1 shows the time evolution of the ambient temperature,
density, pressure, and the heating efficiency, HE, for the
proto-cluster of model 1 of Table 1, for which SFE = 11\%.  After
$\sim 3$ Myr, HE rapidly increases to one. Before that, the total
energy stored in the gas is only a few percent of the total energy
released by the SNe, most of which is radiated away. During this
period, the mass-loss rate from the clouds, mainly due to the
process of photoevaporation (see MGDP), keeps the density of the
diffuse gas high enough allowing an efficient radiative cooling.
After this time, the clouds evaporated almost completely and the
supernovae heating efficiency, HE, increased to unity in less than
0.3 Myr and this leads to a quick gas expansion and removal from
the system. Note that right after 3 Myr the gas temperature rises
to $T\sim 10^7$ K and thus the gas will freely expand through the
cluster radius in $t_{exp} = R/v_s < 0.1$ Myr. These time scales
are  smaller than $\tau_{cr} \sim 0.5$ Myr so that, once HE
reaches the unity value, all the gas will be removed from the
cluster in less than one cluster dynamical crossing time and,
since $\sigma \ge v_{esc}$, or SFE $\le$ 50\%, we may conclude
that the system will become unbound and disperse.

\begin{figure} \centering \epsfxsize=8cm
\epsfbox{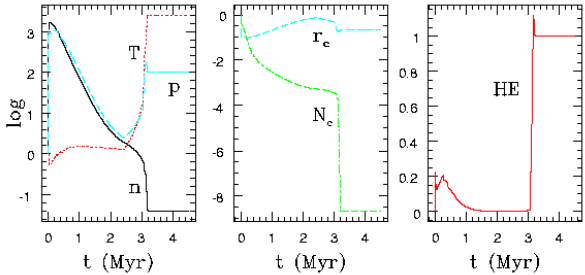} \caption{Ambient density ($n$), temperature ($T$)
and pressure ($p$) evolution in the left panel for a spherical
stellar cluster with initial $M_{star}$ = $10^6$ M$_{\odot}$, R=10
pc, ambient density $n$=10 cm$^{-3}$, $T=10^4$ K, $p = 1.38 \times
10^{-11}$ dy cm$^{-2}$ and total gas mass $M_{gas}=8 \times 10^6
M_{\odot}$. In the center panel: number of clouds ($N_c$) and
clouds radius ($r_c$) evolution; and in the right panel: the SN
heating efficiency (HE).}
\end{figure}

Figure 2a compares HE for systems with different initial radius.
As expected, the larger the radius of the proto-cluster the longer
the system will take to reach a heating efficiency equal to unity,
but in all cases we observe the same trend as in Figure 1, i.e., a
rapid change in HE from $<< 1$ to $\sim 1$ in less than the
corresponding $\tau_{cr} \sim$ 0.5, 1.4, and 2.6 Myr, for the
systems with $R=$ 10, 20, and 30 pc, respectively. Figure 2b shows
the evolution of HE for different SFEs. All the curves have the
same initial conditions
 of Figure 1 except for the initial total gas
mass ($M_{gas}$). We see that the higher the SFE (i.e., the
smaller the total gas content of the system relative to the total
amount of mass in stars) more rapidly the system reaches a high HE
and therefore, earlier is the gas removal. Also, since for the
investigated range of SFE values, SFE = 10\%-50\%, the timescales
for gas removal right after HE reaches the unity value are smaller
than the clusters dynamical crossing times, $\tau_{cr} \sim$ 0.5,
0.8, 1 Myr, for the clusters with  $SFE=$ 11\%, 30\%, and 50\%,
respectively, then they  all should become gravitationally unbound
systems.

\begin{figure}
\centering \epsfxsize=8cm \epsfbox{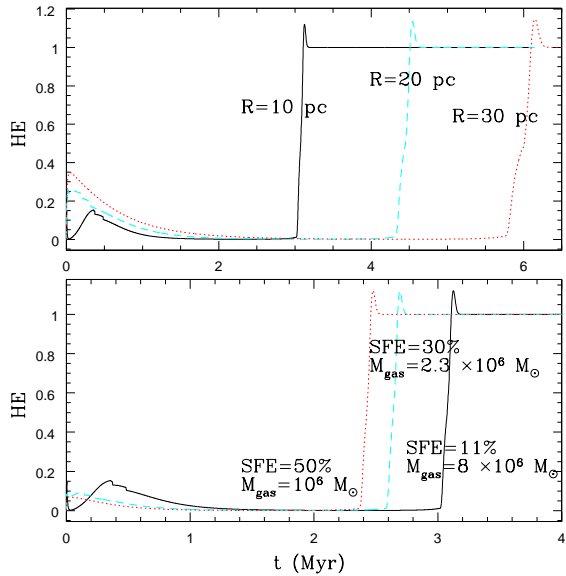} \caption{(a) Top: HE
computed for different initial radius.  All the curves have the
same initial conditions of Model 1 of Figure 1 (with a
$SFE=11\%$), except for the radius of the system. R = 10 pc is
represented by the dot-dashed lines, R = 20 pc by solid lines, and
R = 30 pc by dashed lines. (b) Bottom: Evolution of HE for
different SFEs. All the curves have the same initial conditions of
Model 1 except for the initial gas mass content. SFE = 11\% (with
$M_{gas}= 8 \times 10^6$ M$_{\odot}$) is represented by the
dot-dashed lines, SFE = 30\% (with $M_{gas}= 2.3 \times 10^6$
M$_{\odot}$) by solid lines, and SFE = 50\% (with $M_{gas}= 10^6$
M$_{\odot}$) by dashed lines.}
\end{figure}


Figure 3 adds information to the results of Fig. 2b, showing SFE
versus the time $\tau_{HE}$ that the system takes to reach $HE=1$
for clusters with different initial conditions (Models 1 to 4 of
Table 1). $\tau_{HE}$ can be also interpreted as the time the gas
is retained in the proto-cluster possibly allowing for new
generations of star formation. The curves indicate that this
timescale in a cluster is very insensitive to SFE when the total
initial gas mass content is fixed. $\tau_{HE}$ varies less than
30\% between SFE = 0.1 and 0.5 in each example. Combined with the
results of the previous figures, Figure 3 also reveals that due to
a quick gas removal right after HE becomes equal to one, typical
SB proto-clusters are expected to start to disperse after $\le$ 5
Myr.

\begin{figure}\centering
\epsfxsize=8cm \epsfbox{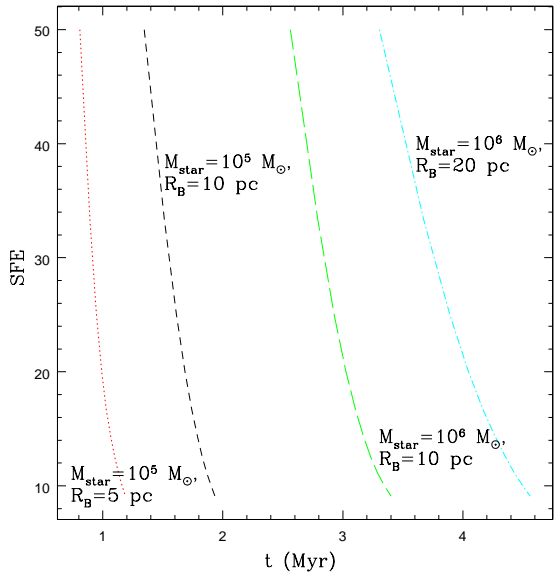} \caption{SFE versus the time
$\tau_{HE}$ that the system takes to reach HE=1 for clusters with
different initial conditions (Models 1 to 4 of Table 1). }
\end{figure}

\section{Conclusions}
We have  investigated  the evolution of the gas in proto-clusters
embedded in  SB systems and have shown that even in the presence
of dense, cold clouds that tend to inhibit the heating of the
interstellar gas by the SNE, the timescales during which the gas
is retained in these systems do not exceed $\sim$ 5 Myr
(considering proto-clusters with total mass in the range
$M_{star}+ M_{gas} \sim 6 \times 10^6 - 6 \times 10^7$ M$_{\odot}$
and radii 5 pc $\le R \le$ 20 pc).  We have also found that after
this time the gas is removed very rapidly ($\sim 0.5$ Myr) in
timescales smaller than the clusters dynamical crossing time and
this result is approximately insensitive to the initial SFE for a
given total gas mass. These very short timescales for gas removal
combined with values of SFE $\le$ 50\% will make these systems to
start to disperse after $\sim$5 Myr.

\begin{acknowledgements}
C.M. and E.M.G.D.P acknowledge financial support from the
Brazilian Agencies FAPESP and CNPq.
\end{acknowledgements}

{}

\label{lastpage}

\end{document}